\documentclass[sigconf]{acmart}

\usepackage{booktabs} 
\renewcommand\footnotetextcopyrightpermission[1]{}
\usepackage{multirow}
\usepackage{subcaption}
\usepackage{caption}
\usepackage{setspace}
\usepackage{amsmath}
\usepackage[english]{babel}
\definecolor{mygray}{gray}{0.5}
\usepackage{float}
\usepackage{flushend}
\usepackage{mathtools}
\usepackage{soul}
\usepackage{arydshln}            
\usepackage{breqn}
\usepackage{enumitem}

\long\def\comment#1{}
\long\def\comments#1{}

\author{Yifan Qiao}
\affiliation{%
  \institution{Department of Computer Science, University of California}
  \city{Santa Barbara}
  \state{California}
  \postcode{93106}
  \country{USA}
}

\author{Shanxiu He}
\affiliation{%
  \institution{Department of Computer Science, University of California}
  \city{Santa Barbara}
  \state{California}
  \postcode{93106}
  \country{USA}
}

\author{Yingrui Yang}
\affiliation{%
  \institution{Department of Computer Science, University of California}
  \city{Santa Barbara}
  \state{California}
  \postcode{93106}
  \country{USA}
}

\author{Parker Carlson}
\affiliation{%
  \institution{Department of Computer Science, University of California}
  \city{Santa Barbara}
  \state{California}
  \postcode{93106}
  \country{USA}
}

\author{Tao Yang}
\affiliation{%
  \institution{Department of Computer Science, University of California}
  \city{Santa Barbara}
  \state{California}
  \postcode{93106}
  \country{USA}
}

\renewcommand{\shortauthors}{Yifan Qiao, Shanxiu He, Yingrui Yang, Parker Carlson \& Tao Yang}



\long\def\comment#1{}
\long\def\comments#1{}

\setcopyright{none} 

\AtBeginDocument{}

\begin{document}

\title{Approximate Cluster-Based Sparse Document Retrieval \\ with Segmented Maximum Term Weights}

\begin{abstract}

This paper revisits cluster-based retrieval that partitions the inverted index into multiple groups
and skips the index partially at cluster and document levels during online inference using a learned sparse representation.
It proposes an approximate search scheme with two parameters to control the rank-safeness competitiveness of 
pruning with segmented maximum term weights within each cluster. 
Cluster-level  maximum weight segmentation allows an improvement in  the rank score bound estimation 
and threshold-based pruning to be approximately adaptive to bound estimation tightness, resulting in better relevance and efficiency.
The experiments with MS MARCO passage ranking and BEIR datasets demonstrate the usefulness of the proposed scheme with a comparison to the baselines.
This paper presents  the design of this approximate retrieval  scheme with  rank-safeness analysis, compares  
clustering and segmentation options, and reports evaluation results.


\end{abstract}

\maketitle

\section{Introduction}

In last few years,
sparse retrieval~\cite{Zamani2018SNRM,
Dai2020deepct, Mallia2021deepimpact, Lin2021unicoil,2021NAACL-Gao-COIL,
Formal2021SPLADE, Formal2021SPLADEV2,Formal_etal_SIGIR2022_splade++, shen2023lexmae}
has exploited transformer-based neural models to learn document tokens with neural weights.
Learned sparse retrieval methods can be effective in relevance on average for in-domain and out-of-domain search  compared  to 
BERT-based dense retrieval with dual encoders (e.g. ~\cite{Ren2021RocketQAv2,Santhanam2021ColBERTv2,Liu2022RetroMAE}).
Its query processing 
is relatively cheap and fast  without GPU support, by taking advantages of  sparse  inverted indices,
specially with index and  model  
optimization ~\cite{2022EMNLP-mackenzie-clipping, lassance2022efficiency,mallia2022faster,20232GT,2023SIGIR-Qiao, 2023SIGIR-SPLADE-pruning}.

Traditional optimization to speed up top-$k$ retrieval given an inverted index is through dynamic pruning techniques,
such as MaxScore~\cite{Turtle1995}, WAND~\cite{WAND}, and BlockMax WAND (BMW)~\cite{BMW}
which  safely skip the evaluation of low-scoring documents that are unable to appear in the final top-$k$ results.
Anytime Ranking by Mackenzie et al.~\cite{2022ACMTransAnytime}
extends the  earlier cluster-based retrieval studies~\cite{2017ECIR-SelectiveSearch,2004InfoJ-ClusterRetr}
to organize the inverted index as clusters with ranges of document IDs,
and  dynamically prune clusters under a time budget.
Such early search termination with a budget can be viewed as a  category of ``rank-unsafe'' pruning. 
There is another line of work on dynamic unsafe pruning called threshold 
over-estimation~\cite{2012SIGIR-SafeThreshold-Macdonald, 2013WSDM-SafeThreshold-Macdonald, 2017WSDM-DAAT-SAAT} and it is
a focus of this paper.

Considering the above advancement  of sparse retrieval with learned representations, 
this paper  revisits  dynamic  index pruning in both safe and  unsafe aspects in the context of cluster-based retrieval. 
Our approach combines cluster-based retrieval with threshold over-estimation and improves its pruning safeness for better relevance effectiveness
or being faster  while retaining the same level of relevance. 
Specifically, the contributions of this paper are listed as follows:

\begin{itemize} [leftmargin=*]
\item 
We propose cluster-level  maximum weight segmentation that partitions a cluster 
into segments offline and  collects segmented term weight information.
This allows cluster-based retrieval to tighten  rank score bound estimation and accurately 
skip more  clusters that only contains low-scoring documents. 


\item We analyze the rank-safeness approximation property of threshold over-estimation and 
propose a two-parameter pruning control to provide  a probabilistic approximate rank-safeness  guarantee. 
This control utilizes   randomly-segmented maximum term weights, and  
makes  cluster-level pruning  approximately adaptive  to  the bound estimation tightness for improved safeness. 

\item
We exploit the use of dense token embeddings as counterparts of sparse vectors under the SPLADE-like models~\cite{Formal2021SPLADE,shen2023lexmae} 
 to cluster documents with k-means and  support the above scheme. 
\item We conduct extensive  evaluations  on    the MS MARCO and BEIR datasets to validate the effectiveness of 
the proposed scheme, called {\bf ASC}. 
This paper  demonstrates  the use of ASC with three  learned sparse retrieval models:
SPLADE~\cite{Formal2021SPLADE,Formal_etal_SIGIR2022_splade++}, uniCOIL~\cite{Lin2021unicoil,2021NAACL-Gao-COIL}, and LexMAE~\cite{shen2023lexmae}.
We  study  its benefit when used with   
two orthogonal efficiency  optimization techniques:  anytime early termination~\cite{2022ACMTransAnytime}
and static index pruning~\cite{2023SIGIR-Qiao}.
\end{itemize}

\comments{
\item To improve recall degradation caused by retrieval approximation,  we compute similarity to top clusters
to give a second chance for revisiting some skipped clusters.
}

\section{Background and Related Work}
\label{sect:background}


{\bf Problem definition.}
Sparse document retrieval identifies top-$k$ ranked candidates that  match a query.
Each document in a data collection is modeled as  a sparse vector with many zero entries. 
These candidates are ranked using a  
simple additive formula, and  the rank score of each document $d$ is defined as:
\begin{equation}
\label{eq:rankscore}
RankScore(d) =\sum_{t \in Q} w_{t,d},
\end{equation}
where $Q$ is the set  of search terms in the given query,
$w_{t,d}$ is a weight contribution of  term $t$  in document $d$, possibly scaled by a corresponding query term weight. 
Term weights can be based on a lexical model  such as BM25~\cite{Jones2000}
or are learned from a neural model such as   DeepImpact~\cite{Mallia2021deepimpact},
uniCOIL~\cite{Lin2021unicoil,2021NAACL-Gao-COIL},
SPLADE~\cite{Formal2021SPLADE}, and LexMAE~\cite{shen2023lexmae}.
Terms are tokens in these neural models.
For a sparse representation, a retrieval algorithm  often  uses 
an \textit{inverted index} with a set of terms, and a \textit{document posting list} for each term.
A posting record in this  list  contains a document ID and  its weight for the corresponding term.

{\bf Threshold-based skipping.} During the traversal of posting lists in document retrieval,
previous studies have advocated for dynamic pruning strategies 
to skip low-scoring documents which cannot appear on the final top-$k$ list~\cite{WAND,strohman2007efficient}.
\comments{
Some information of a posting block $p$
can be accessible without decompression, and such information contains
the maximum weighted term feature score among all documents
and the maximum document ID in this block, denoted  as $BlockMax(p)$ and  $MaxDocID(p)$.
}
To skip the scoring of a document, a pruning  strategy computes the upper bound rank score of a candidate document $d$,
referred to as $Bound(d)$. 
Namely, this bound value  satisfies $RankScore(d) \le Bound(d)$. 

If $Bound(d) \leq  \theta$, where $\theta$ is the rank score threshold to be in the top-$k$ list, this document can be safely skipped. 
WAND~\cite{WAND} uses the maximum term weights of documents  of each posting list to determine their rank score upper bound,
while BMW~\cite{BMW} and its variants (e.g.~\cite{Mallia2017VBMW}) optimize WAND 
using  block-based maximum weights to compute the upper bounds. MaxScore~\cite{Turtle1995} uses  a similar skipping strategy with
term partitioning.
A  retrieval method is  called {\em rank-safe} if it  guarantees that the top-$k$ documents returned are the $k$ highest scoring documents.
All of the above algorithms are rank-safe.
The underlying retriever 
in our evaluation uses MaxScore because it  has been shown  to be more effective for relatively longer queries~\cite{2019ECIRMallia,20232GT}.
That fits in the case of SPLADE, which generates extra query tokens. For example, it has an average of over 23 tokens per query in the MS MARCO Dev set.  

Previous work has also pursued  a ``rank-unsafe'' skipping strategy that  deliberately over-estimates the current top-$k$ threshold 
by a factor~\cite{WAND, 2012SIGIR-SafeThreshold-Macdonald, 2013WSDM-SafeThreshold-Macdonald, 2017WSDM-DAAT-SAAT}.
The previous work on threshold over-estimation does not
have  a formal analysis of  its rank-safeness approximation, whereas  our work provides such an analysis in Section~\ref{sect:property} while
also improving  this technique  for better rank-safeness control in  cluster-based retrieval. 

{\bf Cluster-based document retrieval and  time budgets.}
There is a large body of studies on cluster-based document retrieval  in traditional IR
(e.g.  ~\cite{liu2004cluster,2022SIGIR-KurlandClusterRank,kurland2008opposite})
for re-ranking an initially retrieved list by creating clusters of similar  documents in this list. 
Then the clusters are ranked, and the cluster ranking may be further transformed to a document ranking.  

A {\em cluster skipping inverted index}~\cite{2004InfoJ-ClusterRetr,2017ECIR-SelectiveSearch}
arranges each posting list as  a set of related “clusters”, and those clusters are selected to conduct selective retrieval. 
Anytime Ranking~\cite{2022ACMTransAnytime} extends the above cluster-based retrieval  to selected top clusters
using bound estimations and search under a time budget~\cite{2015ICTIR-anytime-ranking-Lin}. 

\comments{
 as:  
\[
Bound(d) =  \sum_{t \in Q} ClusterMax(t) 
\]
where $ClusterMax(t)$ is the maximum
weight  contribution of term $t$  from all documents in a document range from this cluster.
}
Our work follows the cluster structure design in ~\cite{2022ACMTransAnytime,2004InfoJ-ClusterRetr,2017ECIR-SelectiveSearch},
while increasing index-skipping opportunities through cluster-level maximum weight segmentation and a
probabilistic approximate rank-safeness assurance with a  small impact to relevance.
Both our work and Anytime Ranking belong to   the category of   unsafe pruning, but the key difference is that Anytime Ranking~\cite{2022ACMTransAnytime}
 is focused more on a time-budget driven  
early termination that prunes low-scoring clusters  after sorting with cluster rank score upper bounds.
This  paper  is focused on the  use of threshold over-estimation to prune low-scoring clusters judiciously.
This effort can be combined with Anytime Ranking~\cite{2022ACMTransAnytime} with and  without time budget.
Our evaluation includes a study on  the benefit of  combining  ASC with  Anytime Ranking's early termination technique under a  time budget.  

{\bf Efficiency optimization for learned sparse retrieval.}
There are  orthogonal techniques to speedup  retrieval with learned representations.
BM25-guided pruning skips  documents when  traversing a learned sparse index~\cite{mallia2022faster,20232GT} using extra BM25 weights.
Static index pruning  with thresholding~\cite{2023SIGIR-Qiao,2023SIGIR-SPLADE-pruning}
removes  low-scoring  term weights during offline index generation.
An efficient version of SPLADE~\cite{lassance2022efficiency} uses
L1 regularization for query vectors, dual document and query encoders, and language model  middle training. 
Term impact decomposition~\cite{2022EMNLP-mackenzie-clipping} partitions each posting list into two groups with high and low impact weights.
Our work is complementary to the above optimization techniques.

\section{Cluster based Retrieval with Approximation and Segmentation}

\begin{figure}[htbp]
\begin{center}
  \includegraphics[width=0.9\columnwidth]{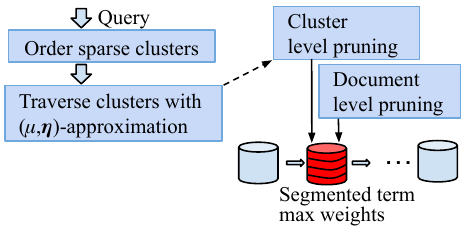}
\end{center}
  \caption{  Flow of ASC: approximate retrieval with segmented cluster-level maximum term weights}
  \label{fig:flow}
\end{figure}

The overall online inference flow of the proposed scheme  during retrieval is shown in Figure~\ref{fig:flow}.
Initially, sparse clusters are sorted in a non-increasing  order of their  estimated cluster upper bounds.
Then, search  traverses the sorted clusters one-by-one  
to conduct approximate retrieval with two-level pruning with segmented term maximum weight. 
Section~\ref{sect:clusterretr} first discusses the use of
 threshold over-estimation  in cluster-based retrieval as $\mu$-approximation. Section~\ref{sect:approximate}
presents $(\mu,\eta)$-approximate search with two-parameter pruning control. 
Section~\ref{sect:property} describes the theoretical properties.
Section~\ref{sect:clustering} discusses the clustering and segmentation options. 

\subsection{Cluster-based $\mu$-approximate retrieval}
\label{sect:clusterretr}

Cluster-based retrieval in this paper follows the concept and data structure
 in~\cite{2022ACMTransAnytime,2004InfoJ-ClusterRetr,2017ECIR-SelectiveSearch}.
A document collection is divided into $m$ clusters $\{C_1, \cdots, C_m \}$.  Then, each posting list of an inverted index is divided following the cluster mapping.

Given a query $Q$, the maximum rank score of a document in a cluster is estimated using the following $BoundSum$ formula and  
Anytime Ranking~\cite{2022ACMTransAnytime} visits sparse clusters in a non-increasing order of $BoundSum$ values.
\begin{equation}
\label{eq:clusterbound}
	BoundSum(C_i) = \sum_{t \in Q}  \max_{d \in C_i} w_{t,d}.
\end{equation}

The visitation to cluster $C_i$ can be pruned if $ BoundSum(C_i) \leq \theta $,
where $\theta$ is the current top-$k$ threshold. 
If this cluster is not pruned, then document-level index traversal and skipping can be conducted within each cluster following a standard retrieval algorithm.
Any document within such a cluster  may be skipped for evaluation if $Bound(d) \leq \theta $ where $Bound(d)$ is computed on the fly based on an underlying retrieval
algorithm such as MaxScore.
 
\begin{figure}[htbp]
\begin{center}
  \includegraphics[width=0.75\columnwidth,trim={1em 1em 1em 2em},clip]{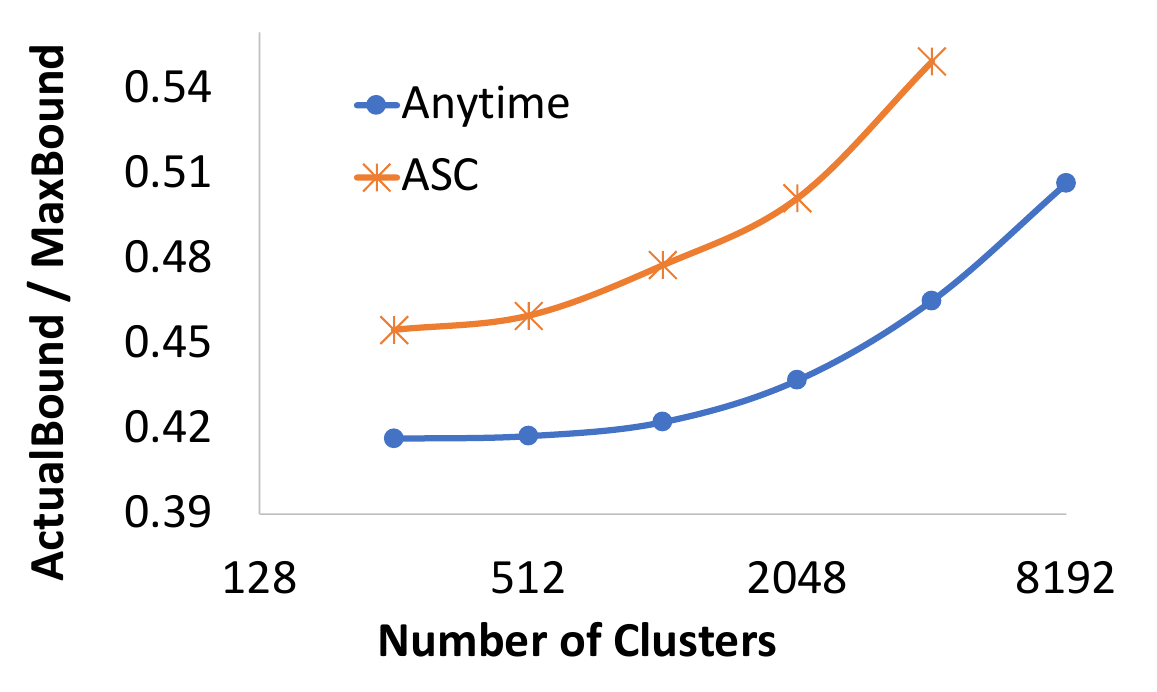}
\end{center}
\caption{The average ratio of the actual and estimated cluster bounds with Formula~(\ref{eq:clusterbound}) on MS MARCO passages}
\label{fig:bounderror}
\end{figure}

The cluster-level bound sum estimation in Formula (\ref{eq:clusterbound}) can be loose, 
especially when a cluster contains diverse document vectors, and this reduces the effectiveness of pruning.
As an illustration, 
Figure~\ref{fig:bounderror} shows the average actual and estimated bound ratio 
using Formula (\ref{eq:clusterbound}) for MS MARCO passage clusters, which is
$
 \frac{1}{m}  \sum_{i=1}^m  \frac{ \max_{d_j \in C_i} RankScore(d_j)}{BoundSum(C_i)}
$, 
where $m$ is the number of clusters.
This ratio with value 1 means the bound estimation is accurate, and a small ratio value  towards 0  means a loose estimation. 
This average  ratio becomes  bigger  with a smaller error when $m$  increases with a smaller  average cluster size. 
This figure  also plots the improved cluster upper bound computed in ASC described below.

Limited threshold over-estimation ~\cite{2012SIGIR-SafeThreshold-Macdonald, 2013WSDM-SafeThreshold-Macdonald, 2017WSDM-DAAT-SAAT}
can be helpful to deal with a loose bound estimation.
Specifically, over-estimation of the top $k$ threshold is applied by a factor of $\mu$ where $0<\mu \leq 1$,
and the above pruning condition is modified as $BoundSum(C_i) \leq \frac{\theta}{\mu} $ and $ Bound(d) \leq \frac{\theta}{\mu}$.
The extension of Anytime Ranking combined with this strategy is called {\bf Anytime$^*$} in this paper.

We call a retrieval algorithm {\em $\mu$-approximate} if 
it satisfies that the average rank score of any top $k'$ results produced 
by this algorithm, where $k' <k$, is competitive to that of rank-safe retrieval within a factor of $\mu$.
 We will show Anytime$^*$ is $\mu$-approximate in Proposition~\ref{propsafe1} of Section~\ref{sect:property}. 

\begin{figure}[htbp]
\begin{center}
  \includegraphics[width=0.9\columnwidth,trim={1em 1em 1em 1em},clip]{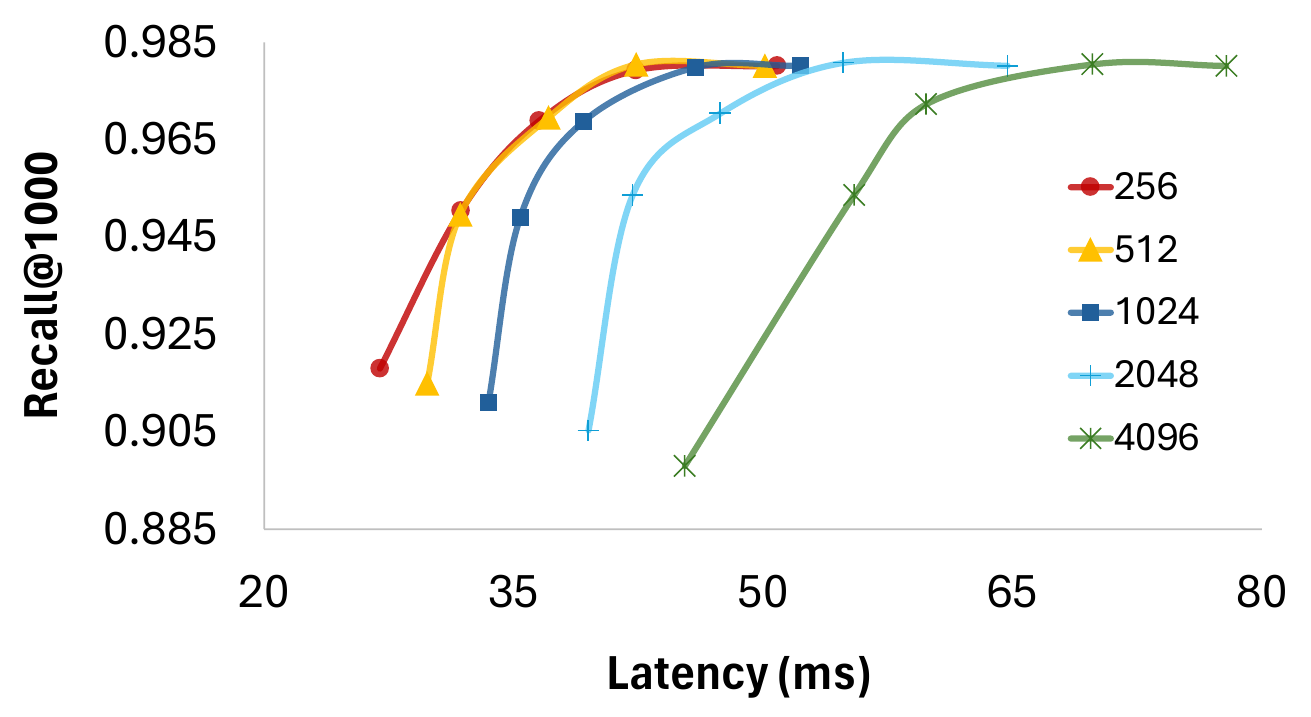}
\end{center}
\caption{Recall vs. latency of Anytime$^*$ on SPLADE with no time budget when $\mu$ varies from 0.3, 0.5, 0.7, 0.9, to  1 for each fixed number of clusters.
Retrieval depth $k=1000$.}
\label{fig:overestimation}
\end{figure}

Figure~\ref{fig:overestimation} shows the relationship of Recall@1000  and latency of Anytime$^*$ on an Intel i7-1260P server
with five different  $\mu$ choices for the MS MARCO Dev set with retrieval depth $k=1000$.  
Each curve in a distinct color represents  a fixed  number of clusters from 256 to 4096.
Each curve contains  five markers from left to right, representing that $\mu$ increases from 0.3, 0.5, 0.7, 0.9, to  1. 
Recall@1000 is retained well when $\mu=0.9$, but drops visibly  as  $\mu$ becomes small.
\comments{ 
The work in ~\cite{2012SIGIR-SafeThreshold-Macdonald, 2013WSDM-SafeThreshold-Macdonald} found $\frac{1}{\mu}$ between  1 and 2 is beneficial while having 
a relevance degradation for BM25 retrieval.  The work in~\cite{qiao2023optimizing} found that even with such a range,
the MRR@10 number of SPLADE++  with $k=10$ drops from 0.3937 to 0.3690 with $\frac{1}{\mu}=1.1$  (namely $\mu=0.91$),
and to 0.321 with $\frac{1}{\mu}=1.3$ for MS MARCO Dev test set.
}
The introduction of   threshold over-estimation with
$\mu$ allows the skipping of more low-scoring documents when the bound estimation is too loose. However,   thresholding is applied uniformly to all cases 
and can incorrectly prune  many desired relevant documents  when the bound estimation is already tight  in some clusters. 
Alternatively, without over-estimation, Anytime Ranking imposes  a time budget to conduct unsafe pruning and  reduce latency.
However, there is no formal approximate rank-safeness guarantee with such an early termination algorithm.


\comments{
{\bf Data needed.}  
\begin{itemize}
\item 1) X-axis: a few settings varying  the number of clusters.\\
Y-axis: the average error sum or squared error between document rank score and cluster boundsum upperbound.

\item 1) X-axis: a few settings varying  the number of clusters.\\
Y-axis: the average error sum or squared error between document rank score and cluster boundsum upperbound.

\item 2) X-axis: a few settings varying  the number of clusters.\\
Y-axis: Latency and relevance of Anytime ranking and Anytime$^*$ with $1/\mu =1.1, 1.3, 1.5$ used in 2GTI paper. Or old 2GTI paper
uses  $1/\mu =1.3, 1.5, 1.7, 1.9$ 
There are too many choses: Let's set $\mu=0.9,0.8$. Then $\mu=0.5$. In our setting, we show we can set $\mu$ as low as 0.5 but maintain $\eta=1$. 
\end{itemize}
}

To improve the tightness of cluster-level bound estimation using Formula (\ref{eq:clusterbound}), 
one can decrease the size of each cluster.  However, there is a significant overhead when increasing the number of clusters. 
One reason is that 
for each cluster, one needs to extract the maximum weights of query terms and estimate the cluster bound, which can become expensive  for
a large number of query terms.
Another reason is that MaxScore~\cite{Turtle1995} identifies a list of essential query terms which  are different  from one cluster to another.
Traversing more clusters yields more overhead for essential term  derivation, in addition to the cluster bound computation.
Figure~\ref{fig:overestimation} shows that, for Anytime$^*$,
as the number of clusters increases,  the aforementioned additional cluster-level overhead can offset the benefit.

 
\subsection{ASC: $(\mu, \eta)$-approximate retrieval with segmented cluster information}
\label{sect:approximate}


The proposed {\bf ASC} method  stands for  
$(\mu, \eta)$-\textbf{A}pproximate retrieval with \textbf{S}egmented \textbf{C}luster-level maximum term  weights.
ASC  segments cluster term maximum weights to  improve the tightness of cluster bound estimation
and guide cluster-level pruning.
It  employs  two parameters, $\mu$ and $\eta$, satisfying $0<\mu \leq \eta \le 1$, 
to detect the cluster bound estimation tightness and improve pruning safeness. 
Details of our algorithm are described below.

\begin{itemize} [leftmargin=*]
\item {\bf Extension  to the cluster-based skipping index.}
Each cluster $C_i$ is subdivided into $n$ segments $\{S_{i,1}, \cdots, S_{i,n}\}$ through random uniform partitioning during offline processing.
The index for each cluster
has an extra data structure which stores the maximum weight contribution of each term from each segment within this cluster.
During retrieval, the maximum and average segment bounds of each cluster $C_i$ are computed as shown below:
\begin{equation}
\label{eq:subclusterbound}
MaxSBound(C_i) = \max_{j=1}^n B_{i,j}, 
\end{equation}
\begin{equation}
AvgSBound(C_i) = \frac{1}{n}\sum_{j=1}^n B_{i,j}, 
\end{equation}
\[\mbox{ and } B_{i,j}  = \sum_{t \in Q}  \max_{d \in S_{i,j}} w_{t,d}.
\]
\item {\bf Two-level pruning conditions}.
Let $\theta$ be the current top-$k$ threshold of retrieval in handling query $Q$.

\begin{itemize} [leftmargin=*]

\item {\bf Cluster-level pruning:}  Any cluster is pruned  when

\begin{equation}
\label{eq:prune1}
   MaxSBound(C_i)  \le \frac{\theta}{\mu} 
\end{equation}
and 
\begin{equation}
\label{eq:prune2}
   AvgSBound(C_i) \le \frac{\theta}{\eta}.  
\end{equation}

\item {\bf Document-level pruning:}
If a cluster is not pruned, then  when visiting such a   cluster with a MaxScore or another retrieval algorithm, 
a document $d$ is pruned if  
\[
 Bound(d)  \le \frac{\theta}{\eta}.
\]

\end{itemize}  
\end{itemize} 

\begin{figure}[htbp]
\begin{center}
  \includegraphics[width=1\columnwidth,trim={1cm 26cm 0 0.5cm},clip]{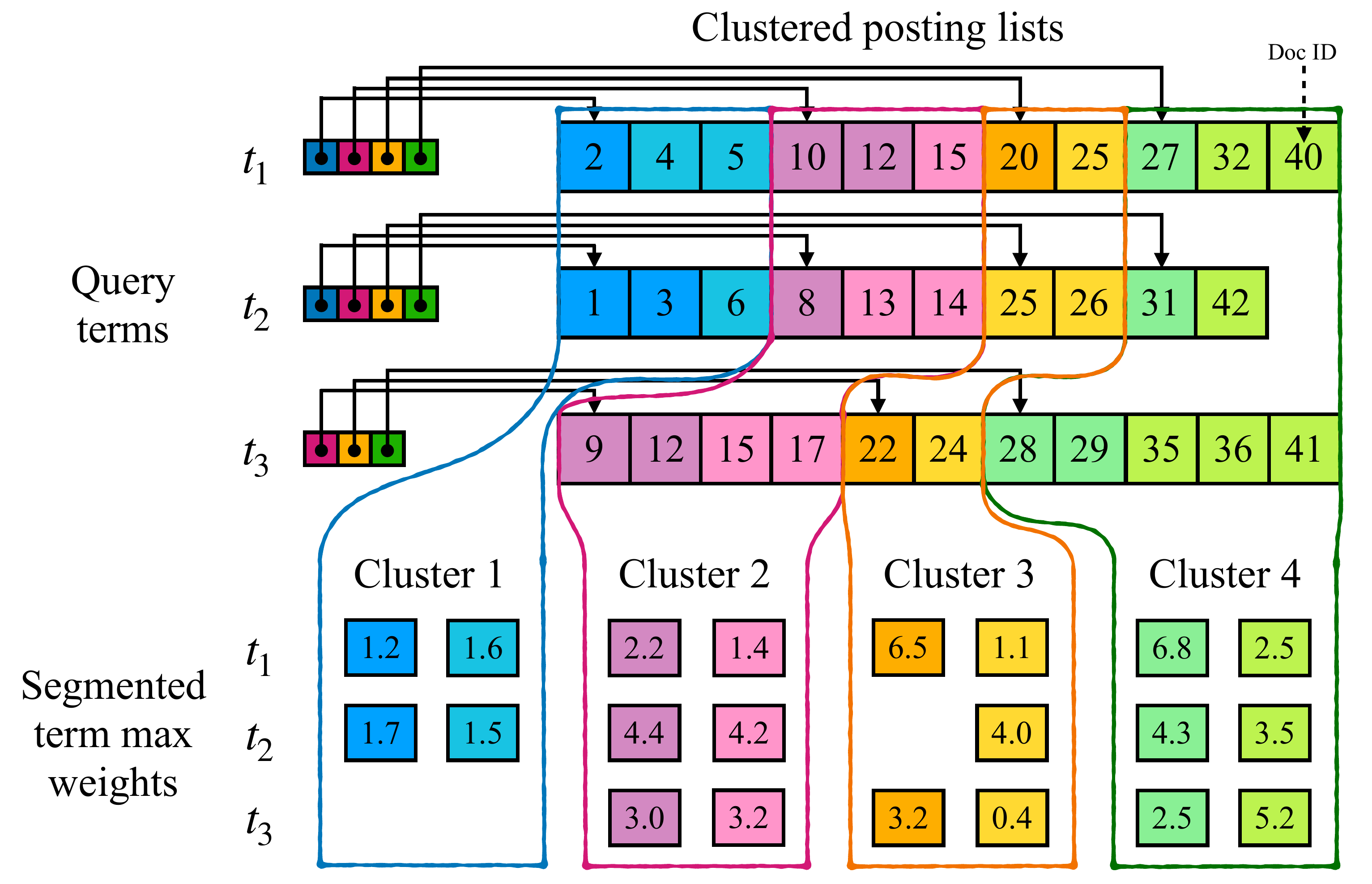}
  \vspace{1em}
{\small (a) Cluster skipping inverted index with 2 weight segments per cluster}
\end{center}
\begin{small}
    \begin{tabular}{|l|c|c|c|c|}
    \hline
	$\theta = 9$ &  {\small Custer 1}    & {\small Cluster 2} & {\small Cluster 3} & {\small Cluster 4} \\ 
    \hline
 $BoundSum$ & 3.3  & 9.8 & 13.7 & 16.3 \\ 
\textbf{\small Anytime} &  {\small Pruned}  & {\small Kept} & {\small Kept} & {\small Kept} \\ 
\textbf{\small Anytime$^*$ $\mu=0.9$} &  {\small Pruned}  & {\small Pruned} & {\small Kept} & {\small Kept} \\ 
    \hline

  $MaxSBound$ & 3.1  & 9.6 & 9.7 & 13.6 \\ 
  $AvgSBound$ & 3.0  & 9.2 & 7.6 & 12.4 \\ 
\textbf{\small ASC } {\small $\mu$=0.9, $\eta$=1 } &  {\small Pruned}  & {\small Kept} & {\small Pruned} & {\small Kept} \\ 
    \hline
    \end{tabular}
    
\vspace{1em}
\end{small}

{\small (b) Decisions of dynamic cluster-level pruning during retrieval} 
  \caption{  A pruning example of Anytime, Anytime$^*$, and ASC}
  \label{fig:clusterindex}

\end{figure}

Figure~\ref{fig:clusterindex}(a)  illustrates a cluster skipping index of four clusters for handling query terms $t_1$, $t_2$, and $t_3$. 
This index is  extended to include two maximum term weight segments per cluster for ASC
and these weights are marked in a different color for
different segments.  Document term weights in posting records are  not shown.
Figure~\ref{fig:clusterindex}(b) lists the cluster-level pruning decision 
in Anytime Ranking, Anytime$^*$, and ASC when the current top-$k$ threshold $\theta$ is 9.  
The derived bound information used for making pruning decisions by these three algorithms is also illustrated. 


{\bf Extra online space cost for segmented maximum weights.}
The extra space cost in ASC 
is to maintain non-zero maximum term weights for multiple segments at each cluster in a sparse   format.
For  example, Figure~\ref{fig:clusterindex} shows four non-zero maximum segment term weights at Cluster 1 are accessed  for the given query. 
To save space, we use the quantized value. Our evaluation uses  1 byte for each weight, which is sufficiently accurate to guide pruning.
\comments{
The maximum extra space for the above  purpose
is $m\times n\times V$ bytes where $m$ is the number of clusters, $n$ is the number of segments per cluster, and $V$ is the average
number of distinct term tokens with non-zero weights in each cluster.
However, the posting list of each term does not necessarily contain documents for each segment, so the actual space used is much lower. 
}
For MS MARCO passages in our evaluation, the
 default configuration has  4096 clusters and 8 segments per cluster. This results in  about 550MB extra space.
With that, the total cluster-based inverted SPLADE index size increases from about 5.6GB for MaxScore without clustering
to 6.2GB for ASC.  This 9\% space overhead is still acceptable in practice. The extra space overhead for Anytime Ranking is smaller because
only cluster-level maximum term weights are needed. 

\subsection{Properties of ASC and Anytime$^*$}
\label{sect:property}

\newtheorem{prop}{{\bf Proposition}}

Table~\ref{tab:property}
compares  the properties of ASC and Anytime$^*$ on their bound relationship and   pruning safeness control in four aspects.
In this context, Anytime$^*$ and ASC are compared without imposing a time budget.
The last row lists the propositions that give a justification accordingly.

We call an algorithm  {\em  $(\mu,\eta)$-approximate } if it is $\mu$-approximate, and
it satisfies that the expected average rank score of any top $k'$ results produced 
by this algorithm, where $k' <k$, is competitive to that of rank-safe retrieval within a factor of $\eta$.
When choosing $\eta=1$, we call a $(\mu,\eta)$-approximate  retrieval algorithm to be {\em probabilistically safe}. 
ASC satisfies the above condition and Proposition~\ref{propsafe2} gives more details.
In our evaluation, the default setting of ASC uses $\eta=1$.

\begin{table}[htbp]
 \small
    \centering
\caption{A comparison of ASC and Anytime$^*$ properties with no time budget} 
\label{tab:property}
\resizebox{1.0\columnwidth}{!}{%
    \begin{tabular}{|l|c|c|c|c|}
    \hline
& \textbf{Cluster} & 
 \textbf{Adaptive to} & 
\textbf{ $\mu$-approx.} &
\textbf{$\eta$-approx.}\\
& \textbf{bound} & 
\textbf{bound quality} & 
\textbf{ safeness} &
\textbf {prob. safeness}\\
    \hline
\textbf{Anytime$^*$} &  Loose    & No & Yes & No \\ 
    \hline
\textbf{ASC} &  Tighter  & Yes & Yes & Yes \\ 
    \hline
\textbf{Proposition} &  ~\ref{prop:tighterbound}    & ~\ref{prop:avgprune}
 & ~\ref{propsafe1} & ~\ref{propsafe2} \\ 
    \hline
     
    \end{tabular}
    }

\end{table}
\begin{small}
\begin{prop}
\label{prop:tighterbound}
\[
BoundSum(C_i) \ge MaxSBound(C_i) \ge \max_{d \in C_i} RankScore(d).
\]
\end{prop}
\end{small}
The above result shows that Formula (\ref{eq:subclusterbound}) provides a tighter upperbound estimation  than Formula 
(\ref{eq:clusterbound}) as demonstrated by   
Figure~\ref{fig:bounderror}.

In ASC, choosing a small $\mu$ value prunes clusters more aggressively, and  
having the extra safeness condition using the average segment bound with $\eta$ counteracts such pruning decisions.
Given the requirement  $\mu \leq \eta$, we can choose $\eta$ to be close to 1 or exactly 1 for being safer. 
When the average segment bound is   close to their maximum  bound in a cluster, 
this cluster may not be  pruned by ASC. This is characterized by the following property.

\begin{prop}
\label{prop:avgprune}
Cluster-level pruning in ASC does not occur to  cluster $C_i$ when one of the two following conditions is true:
\begin{itemize}
\item 
   $MaxSBound(C_i) >  \frac{\theta}{\mu}$ 
\item 
$MaxSBound(C_i) -AvgSBound(C_i) \leq    \left(\frac{1}{\mu} - \frac{1}{ \eta}\right) \theta$.  
\end{itemize}
\end{prop}

\begin{figure}[htbp]
\begin{center}
  \includegraphics[width=0.7\columnwidth,trim={1em 1em 0em 1em},clip]{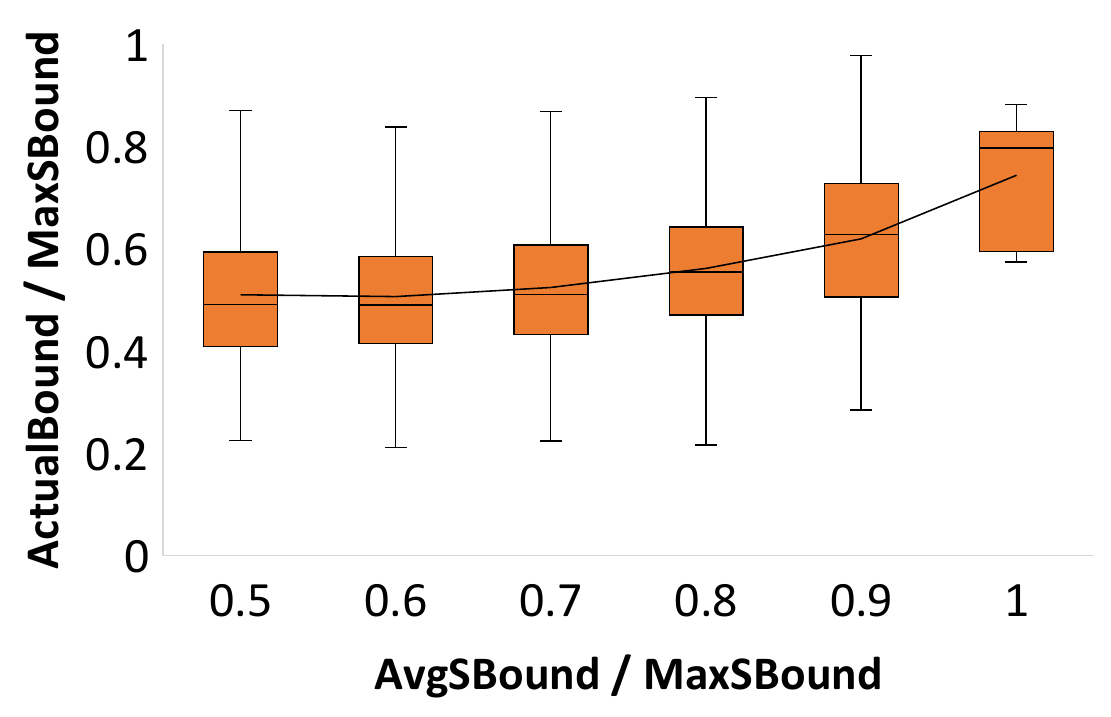}
\caption{
The correlation between bound estimation tightness and  average $AvgSBound(C_i)/MaxSBound(C_i)$ for MS MARCO passage clusters
}
\label{fig:maxavg}
\end{center}
\end{figure}

From the above proposition, when $\mu$ is small and/or the gap between $MaxSBound(C_i)$ and $AvgSBound(C_i)$ is small, 
cluster-level pruning will not occur.  This difference of the maximum and average segment bounds 
provides an approximate indication of the bound estimation tightness with $MaxSBound$,
and Figure~\ref{fig:maxavg} gives an illustration as to why this difference is a meaningful indicator approximately. 
Figure~\ref{fig:maxavg} depicts the correlation between  
the average ratio of  $AvgSBound(C_i)$ over $MaxSBound(C_i)$ for all clusters,
and average ratio of the exact bound over  the estimated bound $MaxSBound(C_i)$. 
The data is collected from the index of MS MARCO dataset with 4096 clusters and 8 segments per cluster. 
This figure shows that when $AvgSBound(C_i)$ is closer to  $MaxSBound(C_i)$ on average, the gap between
exact upper bound  and  $MaxSbound$ value becomes smaller, which means the bound estimation becomes tighter.
Table~\ref{tab:segmentrandom}
in Section~\ref{sect:clustering} will further corroborate  that the above smaller gap yields less cluster skipping opportunities  in ASC
for safer pruning, consistent with the result of Proposition~\ref{prop:avgprune}. 


Define $Avg(x, A)$ as the average rank score of the top-$x$ results   by algorithm $A$.
Let integer $k’ \leq  k$.
The proposition below characterizes  the approximate rank-safeness of pruning in
ASC and Anytime$^*$.
\begin{prop}
\label{propsafe1}
The average top-$k'$ rank score of ASC and Anytime$^*$ without imposing a time budget is
the same as  any rank-safe retrieval algorithm $R$ within a factor of $\mu$.
Namely 
$Avg(k', ASC) \ge \mu Avg(k',  R)$ and
$Avg(k', Anytime^*) \ge \mu Avg(k',  R)$.
\end{prop}

The proposition below characterizes  the extra probabilistic approximate rank-safeness of ASC. 
\begin{prop}
\label{propsafe2}
The average top-$k'$ rank score of ASC achieves the expected value of
any rank-safe retrieval algorithm  $R$  within a factor of $\eta$.
Namely $E[Avg(k', ASC)] \ge \eta  E[Avg(k',  R)]$ where $E[] $ denotes the expected value.
\end{prop}

The probabilistic rank-safeness approximation of ASC relies upon  a condition where each document having  an equal chance 
to be in any segment within a cluster.  That is true because  our segmentation method is  random uniform partitioning.




\subsection{Clustering and segmentation choices}
\label{sect:clustering}

We first discuss how to cluster   documents. 
Here, we assume that  a learned sparse representation is produced from a trained  transformer encoder $T$. For example, 
SPLADE~\cite{Formal2021SPLADE, Formal_etal_SIGIR2022_splade++} and LexMAE~\cite{shen2023lexmae} provide a trained
BERT transformer to encode a document and a query.
There are two approaches to represent documents for clustering:
\comments{
The SPLADE model uses the BERT token space to predict the feature vector $\vec{w}$.
In its latest SPLADE++ model, it first calculates the importance of $i$-th input token in $d$ for
each $j$ in $V$:
$w_{ij}(\Theta)  = {\rm Transform}(\vec{h_i})^T \vec{E_j}+b_j,$
where $\vec{h_i}$ is the BERT embedding of $i$-th token in $d$,
$\vec{E_j}$ is the BERT input embedding for $j$-th token.
Transform() is a linear layer with GeLU activation and LayerNorm.
The weights in this linear layer, $\vec{E_j}$,
 and $b_j$ are the SPLADE parameters updated during training and we call them set  $\Theta$.
Then the $j$-th entry $w_j$  of document $d$ (or a query)  is max-pooled as
$w_j(\Theta) =\max_{i\in d} \{ \log(1+{\rm ReLU}(w_{ij} (\Theta))) \}.$
}

\begin{itemize} [leftmargin=*]
\item {\bf K-means clustering of sparse vectors.} 
Encoder $T$  is applied  to each document in a data collection  
to produce a sparse weighted  vector. 
Similar as Anytime Ranking~\cite{2022ACMTransAnytime},
we follow the approach of ~\cite{2015ACMTrans-KulkarniCluster, 2017InfoRetrJ-KimSelective} 
to apply the  Lloyd's k-means clustering~\cite{k-means}. 
Naively applying the k-means algorithm to the clustering of learned sparse vectors presents a challenge
owing to their high dimensionality and a large number of sparse vectors as the dataset size scales. 
For example, each sparse SPLADE document vector is  of  dimension 30,522 although  most  elements are zero.
Despite its efficacy and widespread use, the k-means algorithm is known to deteriorate when the dimensionality grows.
Previous work on sparse k-means has addressed that  with feature selection and 
dimension reduction~\cite{2020NEURIPS-kmeans, 2020IJCAIkmeans}.
These studies explored dataset sizes much smaller than our context and with different applications. 
Thus our retrieval application demands new considerations. 
Another difficulty is a lack of efficient implementations for sparse k-means in dealing with large datasets. 
We address the above  challenge below by taking advantage of  the dense vector representation produced by the transformer encoder
as counterparts corresponding to their sparse vectors,  with a much smaller dimensionality.  

\item {\bf K-means clustering of dense vector counterparts.} 
Assuming this trained transformer $T$ is BERT, we apply $T$ to each document and produce 
a token embedding set $\{t_1, t_2, \cdots, t_L\}$ and  a CLS token vector.
Here $t_i$ is  the BERT output embedding of $i$-th token in this document
and $L$ is the total number of tokens of this document. Then, we have three ways to produce a dense vector of each document for clustering.
\begin{itemize}
\item The CLS token vector.
\item The element-wise maximum pooling of  all output 
token vectors.  The $i$-th entry of this dense vector is $\max_{j=1}^L t_{i,j}$ where $t_{i,j}$ is the $i$-th entry of $j$-th token embedding.
\item The element-wise mean pooling of all output
 token vectors. The $i$-th entry of this dense vector 
is $\frac{1}{L} \sum_{j=1}^L t_{i,j}$ where $t_{i,j}$ is the $i$-th entry of $j$-th token embedding.
\end{itemize} 
In addition to the above options, we have compared the use of a dense representation 
based on SimLM~\cite{Wang2022SimLM},  a state-of-the-art dense retrieval model. 
 
\end{itemize} 
BERT vectors are of  dimension 768, and   we leverage the FAISS library~\cite{johnson2019billion} 
for dense vector clustering with quantization support, 
which can compress vectors and further reduce the dimensionality.


\begin{table}[htbp]

    \centering
    \caption{K-means clustering of MS MARCO passages  for safe ASC ($\mu=\eta=1$) 
with SPLADE sparse model}
\label{tab:clustering}

    \begin{tabular}{l|cc|cc}
    \hline
     & \multicolumn{2}{c|}{w/o segmentation} & \multicolumn{2}{c}{w/ segmentation} \\
     \textbf{Passage representation} & \textbf{MRT} & \textbf{\%C} & \textbf{MRT} & \textbf{\%C} \\ 
    \hline
    Sparse-SPLADE & 91.6 & 67\% & 70.3 & 53\% \\
    Dense-SPLADE-CLS & 115 & 80\% & 82.7 & 64\% \\
    Dense-SPLADE-Avg & 95.3 & 76\% & 74.2 & 58\% \\
    Dense-SPLADE-Max & 90.8 & 68\% & 71.8 & 54\% \\
    Dense-SimLM-CLS & 105 & 78\%   & 78.5 & 60\% \\
    \hline

    \end{tabular}
\end{table}

Table~\ref{tab:clustering}  compares  the performance of ASC in a safe mode $(\mu=\eta=1)$
using the above five different vector representation options to  apply k-means  clustering
for  SPLADE-based sparse  retrieval.
There are 4096 clusters and eight random segments per cluster.
MRT is the mean retrieval time in milliseconds. 
Column marked with ``\%C''  shows the percentage of clusters that are not pruned during ASC retrieval.
\comments{
We compare the following options with k-means clustering: 
1) dense CLS token vectors generated using the BERT transformer trained in the SPLADE model;
2) average dense  vector  of document tokens transformed  using BERT;
3) sparse document  vectors  encoded using SPLADE.
The above options without segmentation are compared against
}
For sparse vectors, we leverage FAISS dense k-means implementation with sampling, which is still expensive.
Table~\ref{tab:clustering}  
shows that the maximum pooling of SPLADE-based dense token vectors has a similar latency as the sparse vector representation. 
These two options are better than other three options.
Considering the accuracy and implementation challenge in clustering   high-dimension sparse vectors, our evaluation chooses
max-pooled dense vectors derived from the corresponding transformer model.


{\bf Cluster segmentation methods for weight information  collection.} There are two approaches to subdivide each cluster for
segmented term weights   during offline processing. 
\begin{itemize} [leftmargin=*]
\item {\bf Random uniform partitioning}.
This random partitioning
allows ASC to satisfy the probabilistic safeness condition that
each document in a  cluster has an equal chance to appear in any segment.

\item {\bf K-means sub-clustering}.
This is to use the same  clustering for partitioning the entire dataset based on document similarity.
As shown below, this method turns out to be less  effective than random uniform partitioning.
\end{itemize}

\comments{
{\bf Data needed.}
\begin{itemize}
\item 1) X-axis: a few settings varying  the number of clusters or other configurations.\\
Y-axis:  the average error sum or squared error between document rank score and cluster upperbound under different clustering methods.\\
Y-axis: Latency and relevance of our algorithm under different clustering methods.\\
\item 2) Impact on upperbound estimation error, latency, and relevance of using different segmentation: random partitioning vs K-mean subclustering.
\end{itemize}
}

\begin{table}[htbp]

    \centering
        \caption{Segmentation for 4096 clusters (8 segments/cluster)}
\label{tab:segmentrandom}
\resizebox{1.0\columnwidth}{!}{%
    \begin{tabular}{l|cc|cc}
    \hline
    & \multicolumn{2}{|c} \textbf{ K-means subclustering}  & 
    \multicolumn{2}{|c} \textbf{ Random even partitioning}   \\
     \textbf{$\mu$, $\eta$} & \textbf{MRR, Recall} & \textbf{MRT} & \textbf{MRR, Recall} & \textbf{MRT} \\ 
    \hline
    0.3, 1 & 0.3926, 0.9390 & 11.9 & 0.3960, 0.9716 & 20.7 \\
    0.4, 1 & 0.3933, 0.9418 & 12.6 & 0.3960, 0.9718 & 20.8 \\
    0.5, 1 & 0.3954, 0.9588 & 17.7 & 0.3962, 0.9739 & 21.8 \\
    0.6, 1 & 0.3966, 0.9770 & 29.0 & 0.3966, 0.9787 & 27.7\\
    0.7, 1 & 0.3966, 0.9799 & 41.6 & 0.3966. 0.9799 & 38.7\\
    1, 1 & 0.3966, 0.9802 & 69.1 & 0.3966, 0.9802 & 66.6 \\
    \hline
     
    \end{tabular}
    
   }

\vspace{1em}

\begin{small}
    \centering
\label{tab:segmentstats}
\resizebox{1.0\columnwidth}{!}{
    \begin{tabular}{l|c|c}
    \hline
Average ratio	
      & $\frac{Actual}{MaxSBound}$
&$\frac{MaxSbound-AvgSBound}{Actual}$ 
\\ [2pt]
    \hline
    Random even partitioning& 0.55 
&0.49 \\
    K-means subclustering& 0.53 
&0.69 \\
    \hline
     
    \end{tabular}
   }
\end{small}
    
\end{table}

\comments{
1/0.5305-1/0.8378
.69141184588963194976
1/0.5496-1/0.7535
.49236508134286478723
}

The top portion of Table~\ref{tab:segmentrandom} compares the above two segmentation methods
when applying SPLADE to MS MARCO passages with 4098 clusters and 8 segments per cluster.
Random uniform partitioning 
offers equal or better relevance in terms of MRR@10 and Recall@1000, especially when $\mu$ is small.
As $\mu$ affects cluster-level pruning in ASC, random segmentation results in a better prevention of incorrect  aggressive pruning, although 
this can result in less cluster-level pruning and a longer latency. 
To provide an explanation for  the above result, 
the lower portion of Table~\ref{tab:segmentrandom} 
shows the average ratio of actual cluster upper bound over the estimated $MaxSBound$,
and the average difference of $MaxSBound$ and $AvgSBound$ scaled by the actual bound.
Random uniform partitioning provides slightly better cluster upper bound estimation,
while its average difference of $MaxSBound$ and $AvgSBound$ is much smaller than  k-means sub-clustering.
Then, when $\mu$ is small, there are more un-skipped clusters, following Proposition~\ref{prop:avgprune}.
 
The above result also indicates  cluster-level pruning in ASC becomes safer due to its adaptiveness to
the gap between the maximum and average segment bounds, which is  consistent with   Proposition~\ref{prop:avgprune}.
The advantage of random uniform partitioning  shown above corroborates with Proposition~\ref{propsafe2} and
demonstrates the usefulness of   possessing  probabilistic approximate rank-safeness. 

\comments{
{\bf Observations.}
\begin{itemize}
\item K-means partitioning can be faster if $mu$ is set low, since its $AvgSBound$ is tighter, thus more clusters can be pruned. As a result, its relevance drops more when  $mu$ gets smaller.
\item With random partitioning, setting $eta=1$ yields reasonably good relevance no matter how small $mu$ is, reflecting Proposition \ref{propsafe2}.

\end{itemize}
}


\section{Evaluation}
\label{sect:eval}

This section addresses  the following research questions:
RQ1) How does ASC compare against the other baselines, including 
Anytime Ranking and its extension Anytime$^*$,  when no time budget is imposed?
RQ2) What is the benefit of  $(\mu, \eta)$-approximate retrieval  
over  $\mu$-approximate retrieval?
RQ3) Can ASC  be still effective when combined with 
 early  termination of Anytime Ranking with a time budget~\cite{2022ACMTransAnytime} and static index pruning~\cite{2023SIGIR-Qiao}?





{\bf Datasets and metrics.} 
We use the  MS MARCO ranking dataset~\cite{Campos2016MSMARCO,Craswell2020OverviewOT} with 8.8 million passages.  
The average number of WordPiece~\cite{wordpiece} tokens per passage is 67.5.
Following the standard practice in literature,
we report mean reciprocal rank (MRR@10) for  
the development (Dev) query set, which contains  6980 queries,
and report the commonly used nDCG@10 score for 
the TREC deep learning (DL)  2019 and 2020 tracks with 43 and 54 test queries respectively. 
We also report recall, which is the percentage of relevant-labeled results that appear in the final top-$k$ results.
When there are multiple relevance label levels in DL'19 and DL'20, the lowest level is considered as irrelevant, and other levels as relevant.
We test two retrieval depths:  $k=10$ and  $k=1000$.

The second data collection used is  BEIR~\cite{thakur2021beir},
  which contains the 13 publicly available datasets for evaluating zero-shot retrieval performance.
The size of these datasets ranges from 3,633 to 5.4M documents, with the average query length ranging from 5.4 to 193,
and  the average document length ranging from 11.4 to 635.8.


{\bf Experimental setup.} 
The documents are tokenized using the BERT's Word Piece tokenizer. 
We test ASC on a version of SPLADE \cite{Formal2021SPLADE, Formal_etal_SIGIR2022_splade++}, uniCOIL~\cite{Lin2021unicoil,2021NAACL-Gao-COIL}, and LexMAE~\cite{shen2023lexmae}.
For LexMAE, we report two configurations: one encodes passages with title information, which has yielded a higher relevance score than SPLADE for the MS MACRO Dev test set;
the other encodes passages without title information.
Title annotation is  considered to be non-standard in ~\cite{2023SIGIR-Lassance} since the originally released MS MARCO dataset does not utilize this information.
The SPLADE and uniCOIL models do not use title information, following the standard practice. 

\begin{table*}[htbp]

    \centering
        \caption{A comparison with baselines using SPLADE on MS MARCO passages}
\label{tab:msmarco}
    \begin{tabular}{l|c|cc|cc|cc}
    \hline
    
    \hline
    & & \multicolumn{2}{c|}{MS MARCO Dev} & \multicolumn{2}{c|}{DL'19} & \multicolumn{2}{c}{DL'20} \\
     \textbf{Methods} & \textbf{C\%} & \textbf{MRR, Recall} & \textbf{MRT ($P_{99}$)}  & \textbf{nDCG, Recall} & \textbf{MRT} & \textbf{nDCG, Recall} & \textbf{MRT} \\ 
    \hline 
    
    \hline

    \multicolumn{8}{c}{$k=10$. No time budget} \\ 
    \hline


    \multicolumn{8}{l}{{\bf Rank-safe}  }\\
    MaxScore& - & 0.3966, 0.6824 & 26.4 (116) & 0.7398, 0.1764 & 26.3 & 0.7340, 0.2462 & 24.8 \\
     - Anytime Ranking-512 & 69.8\% & 0.3966, 0.6824 & 20.7 (89.3) & 0.7398, 0.1764 & 18.4 & 0.7340, 0.2462 & 17.6 \\
      - ASC-512*16, 
$\mu$=$\eta$=1 
 & 49.1\% & 0.3966, 0.6824 & 15.2 (62.2) & 0.7398, 0.1764 & 15.3 & 0.7340, 0.2462 & 14.8 \\ \hline
    \multicolumn{8}{l}{\bf $\mu$ vs. $(\mu, \eta)$-approximate} \\
    - Anytime$^*$-512-$\mu$=$0.9$ & 62.7\% & 0.3815, 0.6111 & 15.3 (61.1) & 0.7392, 0.1775 & 15.9 & 0.7126, 0.2382 & 15.2 \\
    - ASC-4096*8, $\mu$=$0.9$, $\eta$=$1$ & 7.99\% & 0.3964, 0.6813 & 11.4 (55.9) & 0.7403, 0.1764 & 11.6 & 0.7338, 0.2464 & 11.5 \\  \hline

    \hline
        \multicolumn{8}{c}{$k=1000$. No time budget} \\ \hline

    \multicolumn{8}{l}{{\bf Rank-safe}  }\\
    MaxScore& - & 0.3966, 0.9802 & 65.8 (209) & 0.7398, 0.8207 &  67.0 & 0.7340, 0.8221 & 63.2  \\
    - Anytime Ranking-512 & 93.0\% & 0.3966, 0.9802  & 50.1 (158) & 0.7398, 0.8207 & 54.3 & 0.7340, 0.8221 & 51.1 \\
    - ASC-512*16, 
$\mu$=$\eta$=1 
& 86.3\% & 0.3966, 0.9802 & 45.8 (148) & 0.7398, 0.8207 & 49.9 & 0.7340, 0.8221 & 46.6 \\ \hline
    
    \multicolumn{8}{l}{\bf $\mu$ vs. $(\mu, \eta)$-approximate} \\
    - Anytime$^*$-512-$\mu=0.9$ & 91.4\% & 0.3966, 0.9801 & 46.0 (149) & 0.7398, 0.8205 & 45.1 & 0.7340, 0.8206 & 42.8 \\
    - ASC-4096*8, $\mu$=0.7, $\eta$=$1$ & 21.7\% & 0.3966, 0.9799 & 38.8 (135) & 0.7398, 0.8188 & 40.5 & 0.7340, 0.8218 & 37.3 \\
    \hdashline
    - Anytime$^*$-512-$\mu=0.7$ & 88.9\% & 0.3963, 0.9696  & 37.1 (127) & 0.7398, 0.7881 & 37.9 & 0.7340, 0.7937 & 36.7 \\
    - ASC-4096*8, $\mu$=0.5, $\eta$=$1$ & 8.10\% & 0.3962, 0.9739 & 21.8 (101) & 0.7398, 0.7977 & 22.8 & 0.7355, 0.7989 & 21.7 \\ \hline

    \hline

    \hline
     
    \end{tabular}
\end{table*}

The ASC implementation uses C++, extended from the Anytime Ranking code release~\cite{2022ACMTransAnytime}, which is 
based on the PISA retrieval package~\cite{mallia2019pisa}. 
The inverted index is compressed using  SIMD-BP128~\cite{2015Lemire}, following~\cite{2019ECIRMallia}.
The underlying retrieval algorithm for each cluster is MaxScore.
We applied an efficiency optimization to both ASC and Anytime Ranking code in extracting 
cluster-based term maximum weights when dealing with a large number of clusters. 
All timing results are collected by 
running as a single thread on a Linux server with Intel i7-1260P and 64GB memory.
Our code will be released after publication.



Before timing queries, all compressed posting lists and metadata for tested queries are pre-loaded into memory,
following the same assumption in \cite{khattab2020finding, Mallia2017VBMW}.
For all of our experiments, we perform pairwise t-tests on the relevance between ASC and corresponding baselines. No statistically significant degradation is observed at 95\% confidence level.

\comments{
For the configuration choice of ASC, we will study its  safe pruning setting with $\mu=\eta =1$,
the default setting with $\mu=0.9, \eta=1$ for retrieval depth $k=10$, and $\mu=0.5, \eta=1$ for $k=1000$, and the impact of varying $\mu$.
We will compare ASC with two baselines: Anytime ranking and Anytime$^*$ without time budget imposed.
}

\subsection{Baseline comparison on MS MARCO}


Table~\ref{tab:msmarco} lists the overall comparison of ASC  with two baselines 
using SPLADE sparse passage representations on MS MARCO Dev and TREC DL'19/20 test sets.
Recall@10 and Recall@1000 are reported for retrieval depth $k=10$ and 1000, respectively.  
Retrieval mean response time (MRT) and 99th percentile latency ($P_{99}$) in parentheses are reported in milliseconds.
Column marked ``C\%'' is the percentage of clusters  that are not pruned during retrieval.
For the rank-safe original MaxScore without clustering, we have incorporated a document reordering technique
in \cite{2022ACMTransAnytime} to optimize its index based on document similarity, which shortens its latency by about 10-15\%.

Anytime Ranking and  Anytime$^*$ are configured with 512 clusters and  Anytime$^*$ 
uses  $\mu=0.9$ or 0.7, and these are good choices for the low latency and reasonable relevance based on  Figure~\ref{fig:overestimation}.
ASC is configured with  ``4096*8'' which means 4096 clusters and 8 randomized segments per cluster,
and uses $\eta=1$ with $\mu=0.9$ for $k=10$, and $\mu=0.7$ or 0.5 for $k=1000$. 

\begin{table}[htbp]

    \centering
        \caption{Other learned sparse retrieval models}
\label{tab:sparsemodel}
\resizebox{1.0\columnwidth}{!}{%
    \begin{tabular}{l|cccccc}
    \hline
    
    \hline
      & \multicolumn{2}{c}{uniCOIL} & \multicolumn{2}{c}{\shortstack{LexMAE w/ title}}  & \multicolumn{2}{c}{\shortstack{LexMAE w/o title}}  \\ 
     \textbf{Methods} & \textbf{MRR (R)} & \textbf{T} & \textbf{MRR (R)} & \textbf{T} & \textbf{MRR (R)} & \textbf{MRT} \\
    \hline 
    
    \hline

    \multicolumn{7}{c}{$k=10$. No time budget} \\ 
    \hline


    \multicolumn{7}{l}{{\bf Rank-safe}} \\
    MaxScore& .352 (.617) & 6.0  & .425 (.718) & 47 & .392 (.677) & 46 \\

     - Anytime-512 & .352 (.617) & 5.0 & .425 (.718) & 27 & .392 (.677) & 25 \\
      - ASC-512*16 & .352 (.617) & 4.1 & .425 (.718) & 21 & .392 (.677) & 21 \\ \hline
    \multicolumn{7}{l}{{\bf $\mu$ vs. ($\mu,\eta$)-approximate}} \\
    - Anytime$^*$-$\mu$=0.9 & .345 (.585) & 4.2 & .413 (.654) & 22 & .382 (.623) & 21 \\
    - ASC-$\mu$=0.9,$\eta$=1 & .352 (.614) & 3.9 & .425 (.718) & 16 & .393 (.677) & 15 \\  \hline
    



    \hline
        \multicolumn{7}{c}{$k=1000$. No time budget} \\ \hline

    \multicolumn{7}{l}{{\bf Rank-safe}} \\
    MaxScore& .352 (.958) & 19 & .425 (.988) & 94 & .392 (.983) & 92 \\
    - Anytime-512  & .352 (.958) & 14 & .425 (.988) & 67 & .392 (.983) & 64 \\
    - ASC-512*16 & .352 (.958) & 13 & .425 (.988) & 64 & .392 (.983) & 61 \\ \hline
    \multicolumn{7}{l}{{\bf $\mu$ vs. ($\mu,\eta$)-approximate}} \\
    - Anytime$^*$-$\mu$=0.7 & .351 (.940) & 8.9 & .425 (.978) & 46 & .392 (.972) & 50 \\
    - ASC-$\mu$=0.5, $\eta$=1 & .351 (.946) & 6.4 & .425 (.980) & 26 & .392 (.975) & 25\\ \hline


    \hline

    \hline
     
    \end{tabular}
   }
\end{table}

Comparing the three rank-safe versions for SPLADE in Table~\ref{tab:msmarco}, ASC is about 27\% faster than Anytime Ranking for $k=10$, and 8.6\% faster for $k=1000$,
because segmentation offers a tighter cluster bound as shown in Proposition~\ref{prop:tighterbound}.
For approximate safe configurations 
when $k=10$, ASC 
with $\mu=0.9/\eta=1$ 
has 3.9\% higher MRR@10, 11\% higher recall, and is 25\% faster  than  Anytime$^*$ with $\mu=0.9$. 
When $k=1000$, 
ASC with $\mu=0.7/\eta=1$ is  1.2x faster than Anytime$^*$ with $\mu=0.9$, and
ASC with $\mu=0.5/\eta=1$ is  1.7x faster than Anytime$^*$ with $\mu=0.7$. ASC offers similar relevance scores in both of these two comparisons.
This demonstrates the importance of Proposition~\ref{propsafe2}.
For this reason, ASC is configured to be probabilistically safe with $\eta=1$ while choosing $\mu$ value modestly below 1 for efficiency.
There is a small  relevance degradation compared to the original retrieval, but 
ASC performs competitively while it is up-to 3.0x faster than the original MaxScore without using clusters.
ASC can skip more than 90\%  of 4098 clusters, but its latency does not decrease proportionally compared to the 512-cluster setting.
This is because like Anytime Ranking, increased overhead for dealing  with a large number of  clusters reduces ASC's benefit, as discussed in Section~\ref{sect:clusterretr}.

Table~\ref{tab:sparsemodel} compares ASC with uniCOIL, LexMAE with and without title annotation
in MRR@10, Recall@10 or @1000 (shortened as R), and latency time (shortened as T). 
The conclusions  are similar as the ones obtained above for SPLADE.
\comments{
When applying to SPLADE for MS MARCO Dev set, ASC 
achieves 0.3963  MRR@10 and 0.97 R@1K in 19ms under a probabilistically
safe configuration. In comparison the original SPLADE achieves 0.3966 MRR@10 and 0.9802 R@1K while
taking 78ms (4.1x slower). Anytime Ranking achieves 0.3914 MRR@10 and 0.9543 R@1K in the same time budget.
When applying to LexMAE, ASC2 achieves 0.4239 MRR@10 and 0.976 R@1K in 17.2ms while
the original LexMAE retrieval takes 108ms (6.28x slower) to achieve 0.4252 MRR@10 and 0.9882 R@1K.
}

\subsection{Zero-shot retrieval on BEIR}

     

\begin{table}[htpb]
    \centering
    \caption{Zero-shot performance of SPLADE on BEIR datasets}
            \label{tab:beir}

        \resizebox{1\columnwidth}{!}{%
    \begin{tabular}{l|cr|cr|cr}
    \hline

    \hline
        & \multicolumn{2}{c|}{MaxScore} & \multicolumn{2}{c|}{Anytime$^*$-$\mu=$0.9} & \multicolumn{2}{c}{ASC} \\
\textbf{Dataset} & \textbf{nDCG} & \textbf{MRT} & \textbf{nDCG} & \textbf{MRT} & \textbf{nDCG} & \textbf{MRT}\\
 \hline 

 \hline
 \multicolumn{7}{c}{$k=10$} \\ \hline
 
        DBPedia  & 0.4430 & 81.2 & 0.4314 & 58.1 & 0.4415 & 50.8 \\
        FiQA & 0.3581 & 3.64 & 0.3563 & 2.49 & 0.3584 & 2.67 \\
        NQ  & 0.5551 & 44.9 & 0.5454 & 39.8 & 0.5486 & 25.6  \\
        HotpotQA  & 0.6815 & 323 & 0.6738 & 270 & 0.6798 & 260 \\
        NFCorpus  & 0.3517 & 0.17 & 0.3498 & 0.15 & 0.3516 & 0.17  \\
        T-COVID  & 0.7188 & 5.20 & 0.6727 & 2.48 & 0.7188 & 2.64 \\
        Touche-2020  & 0.3069 & 4.73 & 0.2814 & 2.27 & 0.3069 & 2.00 \\
        ArguAna  & 0.4318 & 9.07 & 0.4110 & 9.17 & 0.4319 & 9.02\\
        C-FEVER  & {0.2429} & 895 & 0.2419 & 735 & 0.2429 & 738  \\
        FEVER   & {0.7855} & 694 & 0.7823 & 587 & 0.7857 & 557  \\
        Quora  & 0.8061 & 5.16 & 0.7949 & 2.05 & 0.8059 & 1.73 \\
        SCIDOCS & 0.1508 & 2.53 & 0.1501 & 2.17 & 0.1509 & 2.13 \\
        SciFact  & 0.6764 & 2.54 & 0.6733 & 2.45 & 0.6764 & 2.42 \\
\hline 
        \textbf{Average } & 0.5007 & - & 0.4896 & 1.43x & 0.5006 & 1.54x \\ \hline

    \hline
 \multicolumn{7}{c}{$k=1000$} \\ \hline
        \textbf{Average } & {0.5007} & - & 0.4982 & 1.96x & 0.4994 & 3.12x \\
 \hline

 \hline
    \end{tabular}
    }
\vspace{-0.8em}
\end{table}


Table~\ref{tab:beir}  
shows the nDCG@10 and mean latency in milliseconds
on zero-shot performance in searching  13 BEIR datasets with retrieval depth $k=10$ and 1000 using SPLADE. For smaller datasets, the number of clusters is proportionally reduced so that each cluster contains approximately 2000 documents, which is aligned with 4096 clusters setup for MS MARCO. The number of segments is kept 8.
ASC uses $\mu=0.9/\eta=1$ for $k=10$ and $\mu=0.5/\eta=1$ for $k=1000$. Anytime$^*$ uses  $\mu=0.9$. There is no time budget imposed.
The  training of the SPLADE model is only based on MS MARCO passages. 

The SPLADE model that we trained has 
an average nDCG@10 score 0.5007, close to the 0.507 reported in the SPLADE++ paper.
ASC offers nDCG@10 similar as MaxScore while being 1.54x faster for $k=10$ and 3.12x faster for $k=1000$. 
Comparing with Anytime$^*$, ASC is 7.7\% faster and has 2.2\% higher nDCG@10 on average for $k=10$, and it is 1.59x faster while maintaining similar relevance scores for $k=1000$.

\subsection{Varying  $\mu$,  \#clusters, and \#segments}

\begin{figure}[h!]
\begin{center}
  \includegraphics[width=0.95\columnwidth,trim={1em 0.5em 2em 1em},clip]{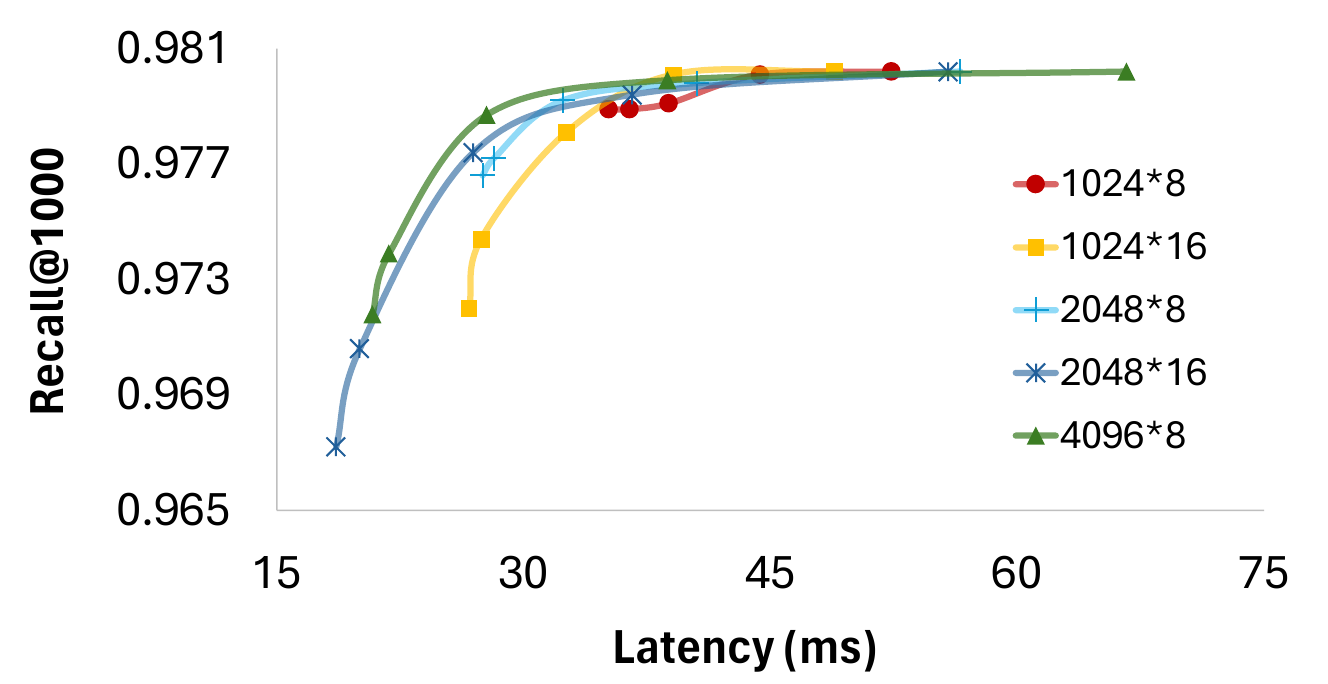}
    \caption{Each curve has different \#clusters and \#segments with  $\mu$ increased from  left to right as 0.4, 0.5, 0.6, 0.7, and 1}
\label{fig:parameterRecall}
\end{center}
\end{figure}

Figure \ref{fig:parameterRecall} examines the relation of Recall@1000 of ASC and  its latency when  varying $\mu$,
the number  of clusters, and the number of segments per cluster on MS MARCO Dev for SPLADE sparse retrieval.
$k=1000$, and Parameter $\eta$ is fixed as 1.
Each curve with a distinct color represents  a different setting   in the number  of clusters and
segments per cluster. ``$m*n$'' means $m$ clusters and $n$ segments per cluster.
Each curve is marked with 5 markers from left to right, representing that $\mu$ increases from 0.3 to  1. 

The curves representing more  clusters have  a longer span, indicating their latency is more sensitive to the value of $\mu$.  
More clusters allows for more accurate cluster bound estimation, and more finer-grained and effective decisions on cluster pruning,
but there is more cluster oriented overhead that affects latency, as discussed in Section~\ref{sect:clusterretr}. 
For safe pruning with $\mu=1$, having 512 clusters is more appropriate for ASC. 
When $\mu<1$, the choice of 4096 clusters  allows for better latency when choosing a small $\mu$ value.

\comments{ 
As the number of clusters gets larger, due to the cluster-based overhead, we observe a leftward shift in the right end of each curve, indicating reduced latency when employing a smaller number of clusters for safe pruning. Consequently, we opt for 512 clusters for secure pruning in both Anytime* and ASC.

A second observation is that a higher number of clusters leads to a longer span for each curve. The reason is, when ASC aims to prune clusters more aggressively when setting a small $\mu$, it can make 
finer-grained and more effective decisions on cluster pruning when each cluster contains relatively small number of documents. Conversely, if the number of clusters is small, each cluster contains more documents, thus the calculated cluster bounds are loose. With $\eta=1$, it becomes challenging to prune an entire large cluster even with a smaller $\mu$. As a result, for unsafe yet effective pruning, a larger number of clusters 4096 is selected.
}



\comments{
Table~\ref{tab:segmentation}  lists the performance of ASC with and without  cluster segmentation.
Starting from safe version of ASC with 4096 clusters, 
ASC with $\mu=1$ and $\eta=1$  and 8 segments per cluster reduces the mean latency from   
90ms to 70.1 with more clusters skipped by safe cluster-level pruning
because of the better accuracy in cluster upper bound estimation based on segmentation.
When $mu=0.5$ and $\eta=1$, the latency drops to only 14ms visiting only 12\% of clusters
and the small degradation in MRR@10 is relatively small
from 0.3966 to 0.3941 while there is a  visible drop in recall@1000.
ASC2 will make up this  loss with a second chance
and boost the relevance  as shown in Table~\ref{tab:graph}. 
The configuration with 1024 clusters shows a similar trend.
}



\comments{
\subsection{Impact of different clustering methods}
\begin{table}[htbp]

    \centering
    \caption{K-means clustering of MS MARCO passages  for safe ASC ($\mu=\eta=1$) 
with SPLADE sparse model}
\label{tab:clustering}

    \begin{tabular}{l|cc|cc}
    \hline
     & \multicolumn{2}{c|}{w/o segmentation} & \multicolumn{2}{c}{w/ segmentation} \\
     \textbf{Passage representation} & \textbf{MRT} & \textbf{\%C} & \textbf{MRT} & \textbf{\%C} \\ 
    \hline
    Sparse-SPLADE & 91.6 & 67\% & 70.3 & 53\% \\
    Dense-SPLADE-CLS & 115 & 80\% & 82.7 & 64\% \\
    Dense-SPLADE-Avg & 95.3 & 76\% & 74.2 & 58\% \\
    Dense-SPLADE-Max & 90.8 & 68\% & 71.8 & 54\% \\
    Dense-SimLM-CLS & 105 & 78\%   & 78.5 & 60\% \\
    \hline

    \end{tabular}
\end{table}

Table~\ref{tab:clustering}  lists the performance of ASC in a safe mode $(\mu=\eta=1)$
with different clustering methods for SPLADE neural retrieval.
We compare the following options with k-means clustering: 
1) dense CLS token vectors generated using the BERT transformer trained in the SPLADE model;
2) average dense  vector  of document tokens transfromered  using BERT; 
3) sparse document  vectors  encoded using SPLADE.
The above options without segmentation are compared against
the maximum pooling of   dense  vectors  of tokens transfromered  using BERT with and without segmentations. 
The result shows that using the maximum pooling for SPLADE-based dense token vectors skips the most clusters and
gives the shortest latency.
That is the reason that this maximum pooling for k-means clustering is used 
in Tables~\ref{tab:msmarco} and \ref{tab:beir}.  
}

\subsection{Compatibility with other efficiency optimization techniques}






     


\begin{table}[h!]

    \centering
        \caption{SPLADE by Anytime and ASC with time budgets}
\label{tab:ortho}
    \begin{tabular}{l|cc}
    \hline

    \hline
     \textbf{Method} &  \textbf{MRR, Recall} & \textbf{MRT ($P_{99}$)} \\ 
    \hline

    \hline
    \multicolumn{3}{c}{$k=10$. Time budget 10ms} \\ \hline
    Anytime-512-$\mu=1$ &  0.3697, 0.6316 & 8.34 (10.3) \\
    ASC-4096*8-$\mu=1, \eta=1$ &  0.3950, 0.6779 & 7.31 (10.1) \\
    \hdashline
    Anytime$^*$-512-$\mu=0.9$ & 0.3604, 0.5750 &  7.70 (10.2) \\
    ASC-4096*8-$\mu=0.9, \eta=1$ & 0.3951, 0.6782 & 6.81 (10.0) \\
    \hline

    \hline
    \multicolumn{3}{c}{$k=1000$. Time budget 20ms} \\ \hline
    Anytime-512-$\mu=1$ & 0.3642, 0.8648 & 19.1 (20.4) \\
    ASC-4096*8-$\mu=1, \eta=1$ & 0.3941, 0.9656 & 19.9 (20.1)\\
    \hdashline
    Anytime$^*$-512-$\mu=0.9$ & 0.3631, 0.8635 & 19.1 (20.3) \\
    ASC-4096*8-$\mu=0.7, \eta=1$ & 0.3954, 0.9695 & 17.0 (20.0) \\
    \hline

    \hline
     
    \end{tabular}
\end{table}

Table~\ref{tab:ortho} lists MRR@10 and  Recall@1000 of combining ASC 
 with early termination technique of Anytime Ranking~\cite{2022ACMTransAnytime} under a time budget 
on MS MARCO Dev set for SPLADE. 
We set time budget 10ms for $k=10$ and time budget 20ms for $k=1000$ respectively.
Comparing with previous results in Table~\ref{tab:msmarco},  
there is a small relevance degradation for ASC with time budgets, but the 99th percentile time 
is improved substantially by this combination. 
Under the same time constraint, this ASC/Anytime combination  has  higher MRR@10 and Recall@1000 
than Anytime Ranking or Anytime* alone in both $k=10$ and 1000.
This result again demonstrates that ASC with $\eta=1$ and $\mu < 1$ is able to filter out some clusters with low ranking scores 
but loose cluster-wide  maximum bounds.


We also apply ASC to a fast version of SPLADE with static
index pruning called HT3~\cite{2023SIGIR-Qiao}. HT3 has 0.3942 MRR@10 on MS MARCO Dev set with a retrieval latency of 24.7ms for retrieval depth $k=1000$. ASC configured with ``4096*8'' and $\mu=0.5/\eta=1$ reduces the retrieval latency by 3.3x to 7.43ms, while the relevance slightly degrades to 0.3933 MRR@10.
            
\section{Concluding Remarks}

This paper has proposed  an approximate sparse retrieval scheme 
with randomly segmented cluster maximum term weights,
and    provided an analysis to characterize the improved rank-safeness 
for better relevance effectiveness. 
The evaluation on the MS MARCO and BEIR datasets has shown the following results.
\comments{ 
The evaluation shows that ACS and ACS2 improve  the previous work and ACS2 is fast  with a small  relevance degradation.
When applying to SPLADE for MS MARCO Dev set, ACS2 
achieves 0.3963  MRR@10 and 0.97 R@1K in 19ms under a probabilistically 
safe configuration. In comparison the original SPLADE achieves 0.3966 MRR@10 and 0.9802 R@1K while
taking 78ms (4.1x slower). Anytime ranking achieves 0.3914 MRR@10 and 0.9543 R@1K in the same time budget.
When applying to LexMAE, ACS2 achieves 0.4239 MRR@10 and 0.976 R@1K in 17.2ms while
the original LexMAE retrieval takes 108ms (6.28x slower) to achieve 0.4252 MRR@10 and 0.9882 R@1K.
}

\begin{itemize} [leftmargin=*]
\item Use of segmented maximum term weights tightens the cluster rank score upper bound estimation, which allows
retrieval to skip more clusters safely.
For example, ASC is 27\% faster for $k=10$ and  8.6\% faster for $k$=1000 than  Anytime Ranking 
using SPLADE and MS MARCO Dev when no time budget is imposed.
\item ($\mu,\eta$)-approximation 
adapts to cluster bound estimation tightness and  adds  probabilistic approximate rank-safeness. 
This  provides  more flexible pruning  control
compared to single-parameter $\mu$-approximation and  allows  ASC to be faster 
than Anytime$^*$ at a similar relevance level or outperform in relevance and efficiency depending on configurations,   when no time budget is imposed. 
For example, with $\mu=0.9/\eta=1$, 
ASC has 3.9\% higher MRR@10, 11\% higher recall, and is 25\% faster  than  Anytime$^*$ with $\mu=0.9$ 
on MS MARCO Dev when $k=10$.  
\item With a time budget, combining  
ASC with Anytime Ranking early termination  has  better  relevance
than  Anytime Ranking or Anytime$^*$ alone in both $k=10$ and 1000. This combination also yields much faster 99th percentile latency.
ASC also works well with a statically-pruned index. 
\item The maximum pooled  dense vectors in a SPLADE-like  model performs competitively to sparse vectors, 
which simplifies the use of  k-means clustering. 
\end{itemize} 
There is a  manageable space overhead for storing cluster-wise segmented maximum weights. 
Increasing the number of clusters for a given dataset is useful to reduce  ASC latency 
up to a point because having more clusters leads to more overhead.

{\bf Acknowledgments}. 
This work is supported in part by NSF IIS-2225942  and has used the computing resource of the ACCESS program 
supported by NSF.
Any opinions, findings, conclusions or recommendations expressed in this material
are those of the authors and do not necessarily reflect the views of the NSF.

\appendix

\section{Proofs of Formal  Properties}

\comments{
{\bf Proof of Lemma \ref{lemma1}.} Let $\theta$ be the top-$k$ threshold at the time visiting this document. Then $\theta \leq \Theta$.
If this document is pruned at the cluster level, then $RankScore(d) \leq  \max_{j=1}^n B_{i,j}  \le \frac{\theta}{\mu}\leq \frac{\Theta}{\mu}$.
If it is pruned at the document level,  $RankScore(d) \leq  \frac{\theta}{\eta} \leq \frac{\theta}{\mu}\leq \frac{\Theta}{\mu}$.
\hfill $\blacksquare$

{\bf Proof of Lemma~\ref{lemma2}. } If document $d$   is pruned at the document level,   $RankScore(d) \leq  \frac{\theta}{\eta}\leq \frac{\Theta}{\eta}$.
If document $d$ is pruned at the cluster level, $ E [RankScore(d)] \leq   \frac{\sum_{j=1}^n B_{i,j}}  {n}\le \frac{\theta}{\eta} \leq \frac{\Theta}{\eta}
$. 
\hfill $\blacksquare$

Define $Avg(x, A)$ as the average score of the top-$x$ results   by algorithm $A$.
Let integer $k’ \leq  k$ and  $S$ be a rank-safe retrieval algorithm. 
}
\comments{
\begin{prop}
The average top-$k'$ rank score of ASC is 
the same as  a safe retrieval algorithm within a factor of $\mu$. 
Namely $Avg(k', ASC) \ge \mu Avg(k',  R)$.
\end{prop}

\begin{prop}
Assume any document $y$  has an equal chance being in any of segments of $y$’s cluster.   
Then the average top-$k'$ rank score of ASC achieves the expected value of  
a safe retrieval algorithm   within a factor of $\eta$.
Namely $Avg(k', ASC) \ge \eta  E(Avg(k',  R))$ where $E[]$ denotes the expected value.
\end{prop}
}

\noindent

{\bf Proof of Proposition~\ref{prop:tighterbound}.}
Without loss of generality, assume in Cluster $C_i$, the maximum cluster bound 
$MaxSBound(C_i)$ 
is the same as the bound of Segment $S_{i,j}$.
Then 
\begin{small}
\[
MaxSBound(C_i) = B_{i,j}  = \sum_{t \in Q}  \max_{d \in S_{i,j}} w_{t,d}
\leq    \sum_{t \in Q}  \max_{d \in C_{i}} w_{t,d} =BoundSum(C_i).
\]
\end{small}
For any document $d$, assume it appears in $j$-th segment of $C_i$, then
\begin{small}
\[
RankScore(d) =\sum_{t \in Q} w_{t,d} \leq \sum_{t \in Q}  \max_{d \in S_{i,j}} w_{t,d}
=B_{i,j} \leq MaxSBound(C_i). 
\]
\end{small}
\hfill $\blacksquare$

{\bf Proof of Proposition~\ref{prop:avgprune}.}
When a cluster $C_i$ is not pruned by ASC, that is because one of 
Inequalities (\ref{eq:prune1}) and (\ref{eq:prune2}) is false.
When Inequality (\ref{eq:prune1}) is true but  Inequality (\ref{eq:prune2}) is false,
we have  
\[
   MaxSBound(C_i)   \leq   \frac{\theta}{\mu}
\mbox{ \ and \ }
    - AvgSBound(C_i) \le  - \frac{\theta}{\eta}.
\] 
Add these two inequalities together, that proves this proposition.
\hfill $\blacksquare$

{\bf Proof of Proposition~\ref{propsafe1}.}
Let $L(x)$ be the top-$k'$ list of Algorithm $x$. 
To prove $Avg(k', ASC) \ge \mu Avg(k',  R)$,
we first remove  any document that  appears in both $L(ASC)$ and $L(R)$
in both side of the above inequality. Then, we only need to show:
\begin{equation*}
\begin{aligned}
 \sum_{d \in L(ASC), d \not\in L(R)}   RankScore(d) 
\ge  \mu * \sum_{d \in L(R), d \not\in  L(ASC)} RankScore(d).
\end{aligned}
\end{equation*}

For the right side of above inequality, if  the rank score of every document $d$ in $L(R)$ (but $d \not\in L(ASC)$) does not exceed
the lowest score
in $L(ASC)$ divided by $\mu$, then the above inequality is true.  There are two cases to prove this condition. 

\begin{itemize} [leftmargin=*]
\item Case 1. If $d$ is not pruned by ASC, then $d$ is ranked below $k'$-th position in ASC.
\item Case 2. Document  $d$ is pruned by ASC when the top-$k$ threshold is $\theta_{ASC}$. 
The final top-$k$ threshold when ASC finishes is $\Theta_{ASC}$. 
If this document $d$ is pruned at the cluster level, 
then $RankScore(d) \leq  \max_{j=1}^n B_{i,j}  \le \frac{\theta_{ASC}}{\mu}\leq \frac{\Theta_{ASC}}{\mu}$.
If it is pruned at the document level,  $RankScore(d) \leq  \frac{\theta_{ASC}}{\eta} \leq \frac{\theta_{ASC}}{\mu}\leq \frac{\Theta_{ASC}}{\mu}$.
\end{itemize}
In both cases, $RankScore(d)$ does not exceed the lowest score in $L(ASC)$ divided by $\mu$.

Anytime$^*$ with no time budget imposed behaves in the same way as  ASC with $\mu=\eta$.
Thus this proposition is also true for Anytime$^*$.

\hfill $\blacksquare$

{\bf Proof of Proposition~\ref{propsafe2}:} 
Define $Top(k', ASC)$ as  the score of top $k'$-th ranked  document produced by ASC. 
$\Theta_{ASC}= Top(k, ASC)$.  

The first part of this proof shows that  for any document $d$ such that $d\in L(R)$ and  $d \not\in L(ASC)$, the following inequality is true:
\[
E[RankScore(d)] \leq  \frac{Top(k', ASC)}{\eta}.
\]
There are two cases that   $d \not\in L(ASC)$:
\begin{itemize} [leftmargin=*]
\item Case 1. If $d$ is not pruned by ASC, then $d$ is ranked below $k'$-th position in ASC.  $RankScore(d) \leq  Top(k', ASC)$.
\item Case 2. 
 If document $d$   is pruned at the document level by ASC when the top $k$-th rank score is $\theta_{ASC}$,
\[   
RankScore(d) \leq  \frac{\theta_{ASC}}{\eta}\leq \frac{Top(k, ASC)}{\eta}
\leq \frac{Top(k', ASC)}{\eta}.
\]

If document $d$ is pruned at the cluster level, 
notice that ASC uses random uniform partitioning, and thus  this
document   has an equal chance being in any segment within its  cluster.   
\[
 E[RankScore(d)] \leq   \frac{\sum_{j=1}^n B_{i,j}}  {n}\le \frac{\theta_{ASC}}{\eta} 
\leq \frac{Top(k, ASC)}{\eta}
\leq \frac{Top(k', ASC)}{\eta}.
\]

\end{itemize}

The second part of this proof shows the probabilistic rank-safeness approximation inequality  based on the expected average top-$k'$ rank score. 
Notice that list size $|L(R)|=|L(ASC)|=k'$, and $|L(R) - L(S) \cap L(ASC) |=|L(ASC) - L(R) \cap L(ASC) |$ where minus notation `-' denotes the set subtraction. 
Using the result of the first part, the following inequality sequence is true:
\begin{equation*}
\begin{aligned}
&E [ \sum_{d \in L(R) } RankScore(d) ]\\
=& E[  \sum_{d \in L(R) \cap L(ASC)} RankScore(d)]
   + E[\sum_{d \in L(R), d \not\in L(ASC)} RankScore(d)]\\
\leq & E[  \sum_{d \in L(R) \cap L(ASC)} RankScore(d)]
   + E[\sum_{d \in L(R), d \not\in L(ASC)} \frac{Top(k', ASC)}{\eta} ]\\
\leq & E[  \sum_{d \in L(R) \cap L(ASC)} RankScore(d)]
   + E[\sum_{d \in L(ASC), d \not\in L(R) }  \frac{RankScore(d)} {\eta}]  \\
\leq &E [ \sum_{d \in L(ASC) }  RankScore(d)]  \frac{1} {\eta}. 
\end{aligned}
\end{equation*}

Thus \ \ 
$ E[Avg(k', ASC)] \ge \eta  E[Avg(k',  R)].$
\hfill $\blacksquare$

\balance
\bibliographystyle{ACM-Reference-Format}
\normalsize
 \bibliography{bib/2024refer.bib,bib/thres.bib,bib/2022extra.bib,bib/2022refer.bib,bib/jinjin_thesis.bib,bib/mise.bib,bib/ranking.bib,bib/reference.bib,bib/url.bib,bib/distill.bib}
\end{document}